# Transfer-printed single photon sources coupled to wire waveguides

AUTHOR NAMES

*Ryota Katsumi[1], Yasutomo Ota[2], Masahiro Kakuda[2], Satoshi Iwamoto[1,2] and Yasuhiko Arakawa[1,2]*

AUTHOR ADDRESS

[1] Institute of Industrial Science, The University of Tokyo, 4-6-1 Komaba, Meguro-ku, Tokyo, Japan

[2] Institute for Nano Quantum Information Electronics, The University of Tokyo, 4-6-1 Komaba, Meguro-ku, Tokyo, Japan

E-mail: katsumi@iis.u-tokyo.ac.jp, ota@iis.u-tokyo.ac.jp, arakawa@iis.u-tokyo.ac.jp

Corresponding author: Yasutomo Ota, Yasuhiko Arakawa

**ABSTRACT**

Photonic integrated circuits (PICs) are attractive platforms to perform large-scale quantum information processing. While highly-functional PICs (e.g. silicon based photonic-circuits) and high-performance single photon sources (SPSs, e.g. compound-semiconductor quantum dots (QDs)) have been independently demonstrated, their combination for single-photon-based applications has still been limited. This is largely due to the complexities of introducing SPSs into existing PIC platforms, which are generally realized with different materials and using distinct fabrication protocols. Here, we report a novel approach to combine SPSs and PICs prepared independently. We employ transfer printing, by which multiple desired SPSs can be integrated in a simple pick-and-place manner with a theoretical waveguide coupling efficiency >99%, fulfilling the demanding requirements of large-scale quantum applications. Experimentally, we demonstrated QD-based SPSs with high waveguide coupling efficiencies, together with the integration of two SPSs into a waveguide. Our approach will accelerate scalable fusion between modern PICs and cutting-edge quantum technologies.

The rapid evolution of photonic integrated circuit (PIC) technologies, as represented by the current flourish of silicon photonics [1], is a basis for modern information technologies. State-of-the-art PICs are being used for an expanding array of applications [2], such as photonics-based artificial neural networks [3] with laser light inputs. Introducing quantum light into PICs [4,5] will allow for advanced PIC-based optical information processing, such as linear optical quantum computation [6].

To that end, it is vital to use near-ideal quantum light sources [7]: for example, it is required that single photon sources (SPSs) provide single photons with near-unity efficiency, indistinguishability and purity. Even after the long-term development of diverse SPS technologies, only a few solid state materials, including InAs/GaAs semiconductor quantum dots (QDs) [8–12], have currently been proven to potentially fulfill these demanding requirements [7]. As such, the combination of such SPSs with existing high-end PIC platforms is an immediate route for the realization of large-scale quantum PICs.

In this context, the hybrid integration of SPSs into PICs is very promising. In previous demonstrations [13,14], photonic structures for the SPSs and the waveguides have been jointly processed on single wafers made, for example, by conventional wafer bonding [13]. Such a joint-fabrication process of hybridized material platforms could hinder the highly optimized fabrication of each element: indeed, experimental waveguide coupling efficiencies of so-far demonstrated hybrid SPSs have been limited to around 10~40%. The necessary complicated process flows will also hamper the straightforward use of existing PIC platforms that are in general fabricated in specially-customized facilities (like complementally-metal-oxide (CMOS) process foundries). Very recently, the hybrid integration of QD SPSs on silicon waveguides by using a micro manipulator has been reported [15], succeeded in the independent preparation of the SPS and the waveguide for high quality fabrication. However, the demonstrated waveguide coupling efficiency is still far less than the unity.

In this Letter, we propose and demonstrate an alternative approach based on transfer printing [16–20], by which SPSs and PICs can be prepared independently and integrated easily in a pick-and-place manner. The SPS is based on a QD in a nanocavity and is placed above a waveguide buried in glass, supporting near-perfect theoretical coupling of single photon emission into the waveguide. Transfer printing largely simplifies the required three-dimensional integration of the optical elements, allowing for the demonstration of QD-based SPSs with high experimental coupling efficiencies as well as the dense integration of two SPSs into a waveguide.

Figure 1 shows the basic flow of the proposed hybrid integration process (see Supplementary Information for more details). First, we prepare a QD wafer (Fig.1(a)) for the fabrication of nanocavity-based air-bridge SPSs (Fig.1(b)). We also use another wafer (Fig.1(d)) to separately fabricate wire waveguides buried in glass cladding (Fig. 1(e)), which, if necessary, can be prepared by CMOS process foundries [21]. Then, we use transfer printing to pick up a suitable SPS from the processed QD substrate using a transparent rubber stamp (Fig.1(c)). We transfer the SPS onto the waveguide under an optical microscope: the SPS can be released on the waveguide by slowly peeling the stamp off (Fig.1(f)). A schematic of the completed waveguide-coupled SPS is shown in Fig. 1(g). This structure enables near-unity coupling of QD emission into the waveguide, as discussed later. By pre-selecting suitable SPS candidates prior to the transfer (with appropriate emitter linewidths, positions and wavelengths), the transfer-printing approach may solve major difficulties for incorporating multiple solid-state SPSs into PICs [7,22].

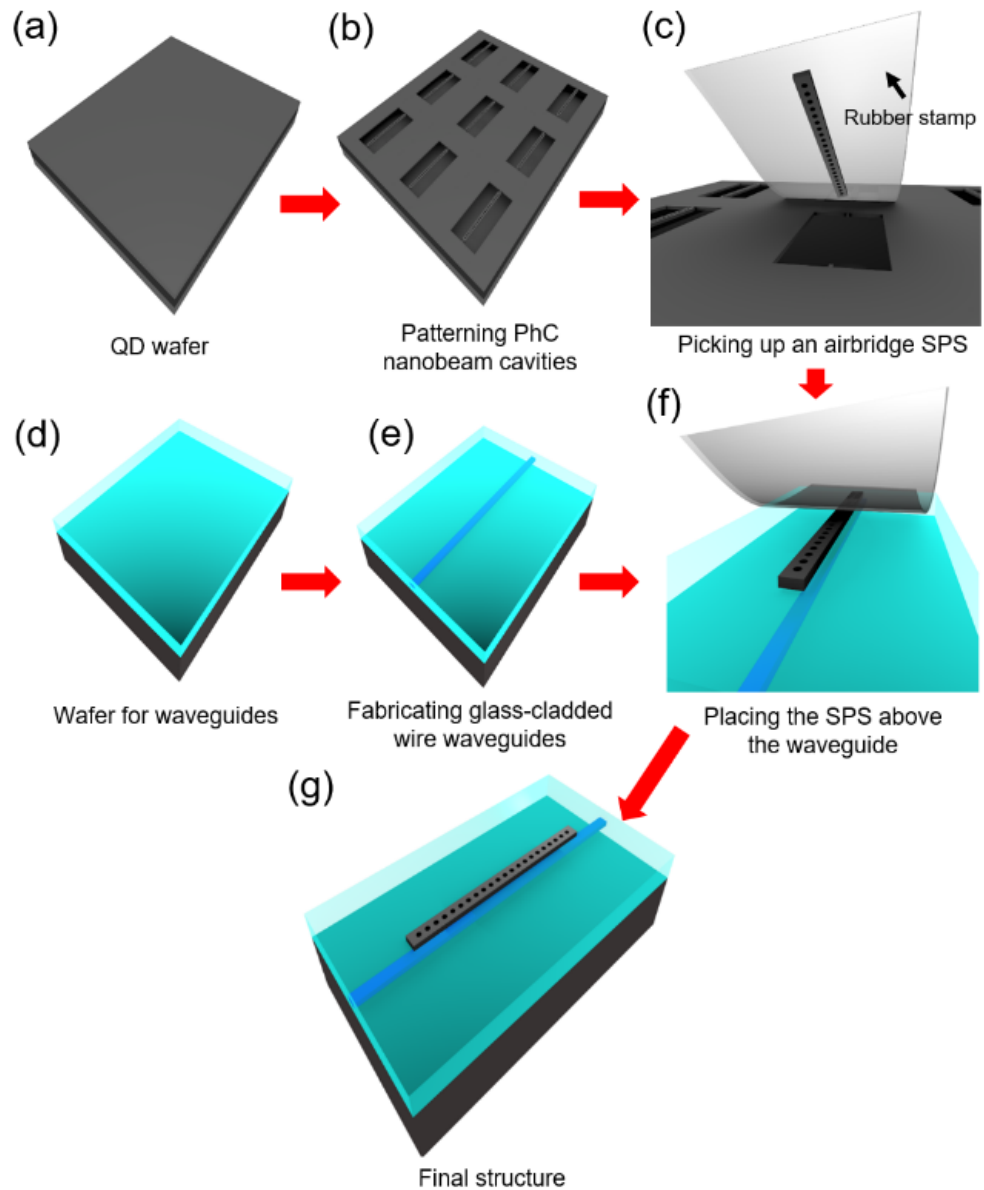

**Fig. 1.** The process begins with the preparation of SPSs on (a) a QD wafer, by patterning (b) an array of PhC nanobeam cavities. (c) An airbridge SPS is picked up by attaching and quickly peeling off the transparent rubber stamp. Meanwhile, we use (d) another wafer for preparing (e) glass-cladded wire waveguides. (f) The picked-up SPS is transferred by placing it above the waveguide and slowly releasing the stamp. (g) Final structure.

Our design of the SPS structure allows near-unity coupling of QD radiation into the waveguide. Figure 2(a) and (b) respectively show top-view and cross sectional schematics of the waveguide-coupled SPS. The nanocavity is based on a one-dimensional photonic crystal (PhC) with local lattice deformation [23–25]. Here we utilize the fundamental cavity mode, which possesses a high Q-factor of $5.4\times10^6$ and a small mode volume of $V = 8.6 \times 10^{-3}$ $\mu m^{-3}$ when located solely on a glass clad without the waveguide. When introducing the waveguide underneath [26], the cavity Q-factor exhibits an exponential dependence on the cavity-waveguide distance, $d$, as shown in Fig. 2(c). The sharp reduction in $Q$ stems predominantly from the coupling of light into the waveguide, characterized with a cavity-waveguide coupling efficiency $\eta$ of over 99% for 250 nm $< d <$ 450 nm. Further reduction of $d$ leads to a degradation of $\eta$, as the waveguide becomes too close to the cavity and starts to scatter cavity photons into free space. In the design presented here, we tuned the cavity/waveguide parameters to minimize the free space scattering for a given $d$ (see Supplementary Information).

The emitter-cavity coupling efficiency, $\beta$, reduces when decreasing cavity $Q$ due to the reduction of the Purcell effect, which scales with $Q/V$. Nevertheless, even when $d = 350$ nm ($Q =$ 3,800), the maximum possible $\beta$ is as large as 99.9%, thanks to the very small $V$. Overall, the theoretical maximum single photon coupling efficiency from the QD into the waveguide, $\eta\beta$, is deduced to be a near-unity value of 99.7% for $d = 350$ nm (see Supplementary Information). We also note that such near-unity $\eta\beta$ can be obtained even in different material combinations (see Supplementary Information).

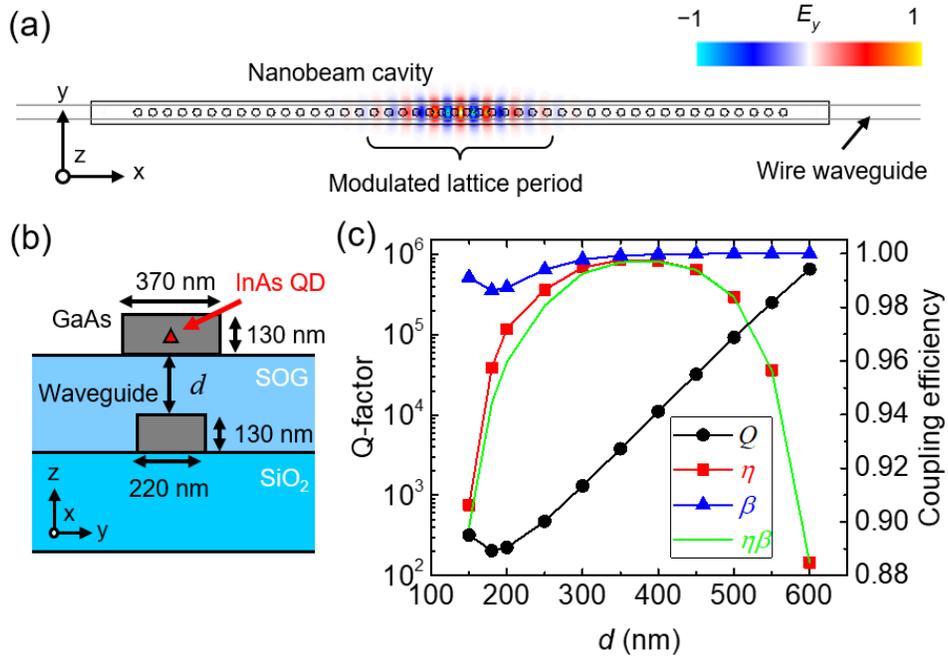

**Fig. 2.** (a) Top view of the designed PhC nanobeam cavity coupled to the waveguide underneath, displayed together with the electric field distribution ($E_y$ component) of the fundamental cavity mode. (b) Schematic cross section of the waveguide-coupled SPS. (c) Simulated cavity Q-factors (black), $\eta$s (red), $\beta$s (blue) and total single photon coupling efficiencies $\eta\beta$s (green), plotted as a function of $d$.

Figure 3(a) shows an optical microscope image of a fabricated sample. For the waveguide, we fabricated a glass-clad GaAs structure using transfer printing and spin-on-glass coating (see Supplementary Information). From the microscope image, a good alignment between the top nanobeam cavity and the underlying waveguide can be seen. Indeed, there is less than 100 nm position deviation between the nanobeam and the waveguide in the *y* direction. The waveguide is terminated by two exit ports, which are composed of diffraction gratings to direct the single photons into free space [27]. In order to optically characterize this structure, we performed low temperature photoluminescence (PL) measurements (see Appendix A). As an initial experiment, we focused a pump laser beam onto the center of the nanocavity and measured a sample image, as displayed in Fig. 3(b). We observed bright light out-coupling from the exit ports, indicating efficient waveguiding of the cavity mode emission.

Then, with low optical pumping to the cavity center, we measured emission spectra of the radiation from one of the exit ports, as shown in Fig. 3(c). In the upper red curve, an intense cavity mode emission at 902.5 nm, together with cavity-coupled QD emission is clearly seen. The measured cavity $Q$ was 3,600, which is 0.28 times smaller than those measured for the nanocavities placed on flat glass. This significant reduction of the Q-factor suggests a large experimental cavity-waveguide coupling efficiency ($\eta_{exp}$) of 72%. The lower, green, spectrum in Fig. 3(c) is of the emission measured above the cavity center. This spectrum does not show the cavity peak, implying that the leakage into free space is largely suppressed, and that the emission occurs predominantly into the waveguide.

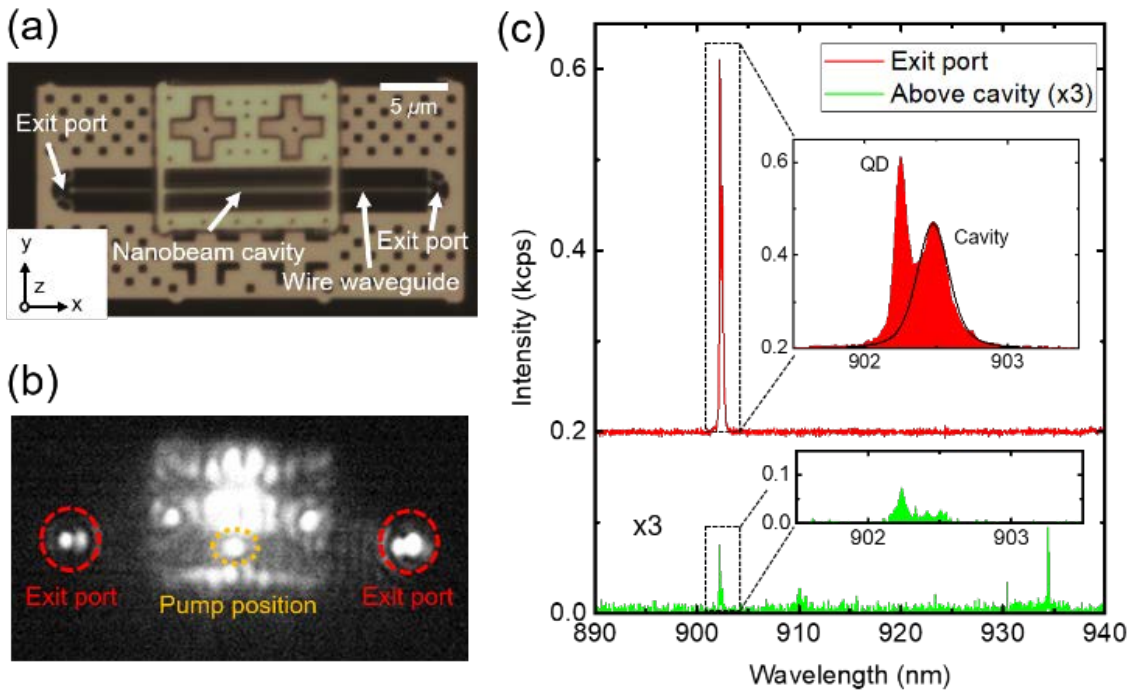

**Fig. 3.** (a) Visible microscope image of a completed device. (b) Low temperature (7 K) PL image of the sample. (c) PL spectra taken by pumping the cavity center with an 815-nm continuous wave laser with a power of 20 nW. The red curve is the spectrum measured through the exit port. An offset is added to the curve. The green curve is measured above the cavity by spatially selecting the radiation near the cavity center.

We performed further detailed optical characterization of the fabricated sample. Figure 4(a) shows a color plot of the temperature dependent emission spectra measured through the exit port. We observed an enhancement of the QD emission near the cavity resonance [10,11], suggesting that the increase of $\beta$ is due to the Purcell effect. We also performed time-resolved PL measurements under the QD-cavity resonance, as shown in Fig. 4(b). A rapid decay of the emission with a rate of 3.83 ns$^{-1}$ is observed, which is 3.8 times faster than that of unprocessed QDs (as shown by the gray curve). The estimated emitter-cavity coupling efficiency from these measurements ($\beta_{exp}$) is 87% (see Supplementary Information). Given $\eta_{exp} = 72\%$, the total single photon coupling efficiency, $\eta_{exp}\beta_{exp}$, is estimated to be 63%.

Next we performed second order correlation measurements based on a Hanbury Brown-Twiss setup when the QD is slightly detuned from the cavity resonance by 0.43 nm. Figure 4(d) shows a measured intensity correlation histogram, exhibiting a clear anti-bunching with a zero delay time value of the second order coherence function of $g^{(2)}[0] = 0.23$, demonstrating single photon generation from the QD.

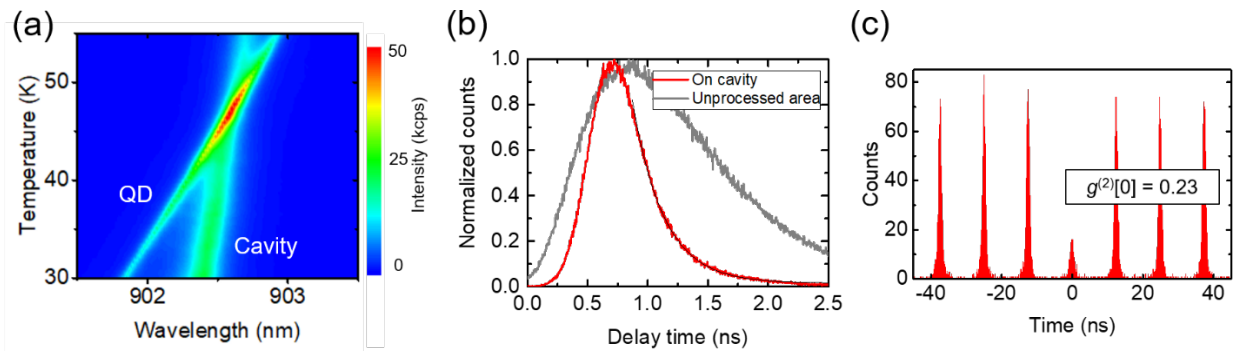

**Fig. 4.** (a) Color plot of temperature dependent emission spectra measured through the exit port. (b) Time-resolved PL spectra measured by time-correlated single photon counting. The red curve was taken at 46 K corresponding to the condition that the QD peak is resonant with the cavity mode. The black curve shows the fitting result, and the gray curve is measured for a typical QD embedded in an unprocessed area of the sample. (c) Measured second order coherence function.

Finally, we extended our method to integrate multiple SPSs into a waveguide, which will be required for the realization of large-scale quantum PICs. For this demonstration we used the same waveguide platform and were able to integrate two different SPSs by repeating the transfer printing process. A microscope image of the completed sample is shown in Fig. 5(a). The nanocavities are designed to resonate at different wavelengths, such that any disturbance on the transport of single photons by the other cavity is largely suppressed (see Supplementary Information). We characterized the optical performance of the two SPSs using PL measurements through the waveguide exit ports, as shown in Figs. 5(b)-(g). For both SPSs, strong QD emission peaks are observed (Figs. 5(b) and (c)). These peaks show fast radiative decay rates as can be seen in time-resolved PL spectra for each emission line (Figs. 5(d) and (e)), indicating their Purcell-enhanced emission into the cavity modes. From these results, we deduced high $\eta_{exp}\beta_{exp}$ efficiencies of 74% and 52% for the left and right SPS, respectively. In addition, we confirmed that the QD emission peaks exhibit strong anti-bunching, as shown in Figs. 5(f) and (g), suggesting the successful integration of two Purcell-enhanced, efficient SPSs into the individual waveguide.

In summary, we have demonstrated the transfer-printing-based integration of QD SPSs into wire waveguides. We have shown that our strategy allows near-unity total single photon coupling efficiencies into the waveguide, while largely relaxing the difficulties in the hybrid integration of SPSs into PICs. Experimentally, we demonstrated high $\eta_{exp}\beta_{exp}$ efficiencies of up to ~70%, which we expect will be further improved by elaborating the nanofabrication (see Supplementary Information). We believe that our approach will be a key technology for the fusions between the state-of-the-art SPSs and modern PICs, irrespective of material choice.

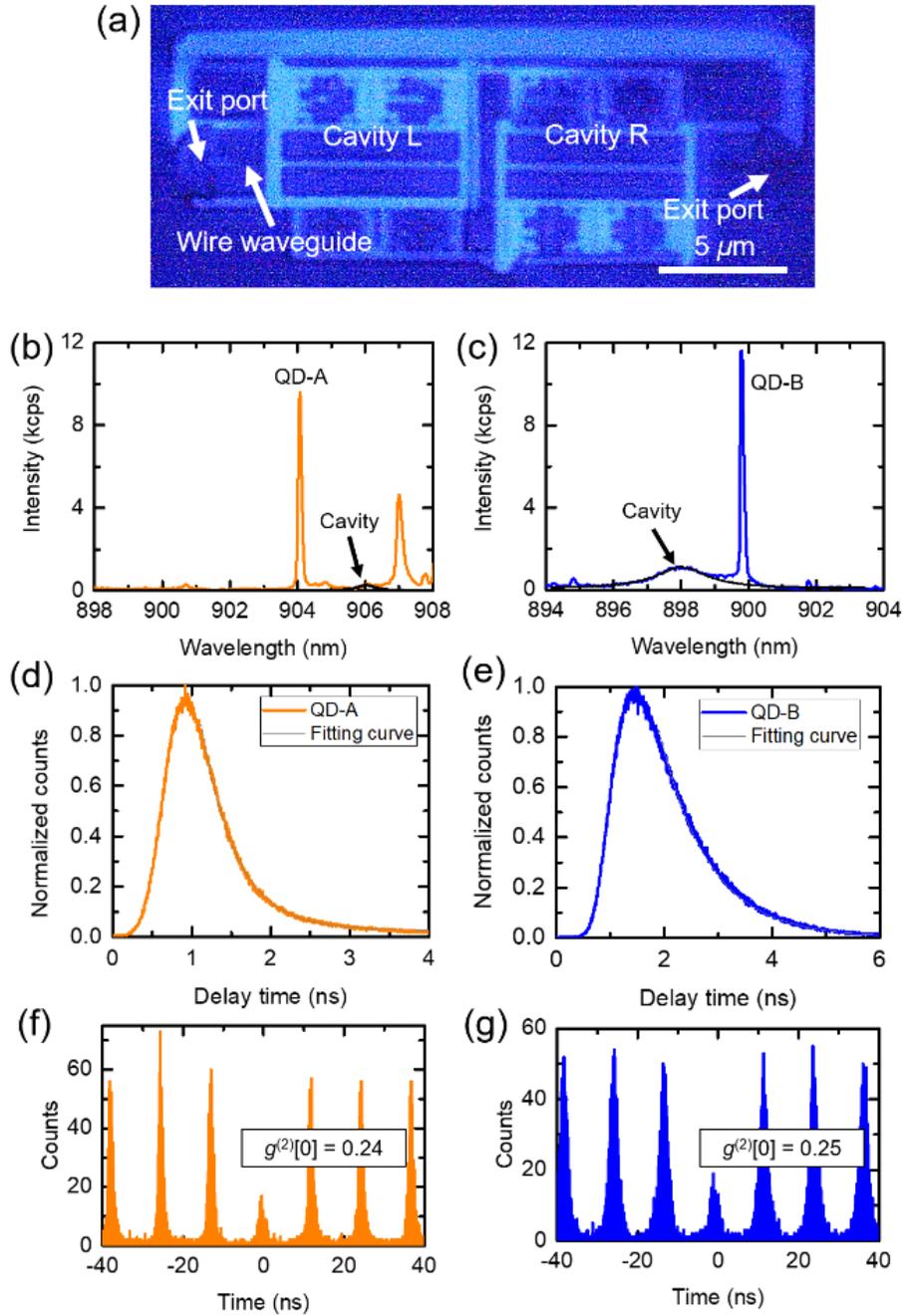

**Fig. 5.** (a) Microscope image of the completed sample. Two nanobeam cavities (cavity L and R) are integrated onto the single waveguide. (b)-(g) Emission properties of the QDs embedded in the (b, d, f) left and (c, e, g) right cavity, respectively. (b) and (c) PL spectra measured through the exit port at 3 K. The labels, QD-A and QD-B, indicate the investigated QD emission peaks. The black solid lines indicate the cavity peak positions. (d) and (e) Time-resolved PL spectra. The black solid lines show fitting results. (f) and (g) Measured second order coherence functions.

**Appendix A: Optical characterization**

The PL measurements were conducted with a low-temperature micro-PL setup. The sample was fixed in a helium flow cryostat having a built-in heater for temperature control. We used an objective lens with a numerical aperture of 0.65 for sample imaging, pump laser beam focusing and collecting PL signal. The collected PL was analyzed with a grating spectrometer equipped with a Si CCD camera. For measuring the PL image in Fig. 3(b), we used a continuous wave titanium sapphire laser oscillating at 819 nm with a pump power of 25 µW (measured before the objective lens), which was focused onto the cavity center. In this experiment, we inserted a bandpass filter centered at 900 nm in front of an imaging camera for extracting the contribution of the cavity mode emission. For measuring the spectra in Fig. 3(c), we switched the pump source to a pulse laser (pulse width = ~ 1 ps, repetition rate = 80.3 MHz) oscillating at 815 nm with an average pump power of 20 nW. We limit the area of PL collection (spatial filtering) around one of the exit ports by narrowing the entrance slit of the spectrometer and the region of interest of the CCD, which roughly corresponds to the detection area of a few µm$^2$. The PL spectra in Fig. 4(a) was measured with the CW laser (wavelength = 849 nm, power = 87 nW). The sample temperature was controlled using the heater and the QD-cavity resonance condition was achieved at 46 K. The results in Fig. 4(b) and (c) were obtained with the pulsed laser (815 nm, 20 nW). The time resolved spectra were measured with a time-correlated single photon counting technique with a silicon avalanche photodiode (overall time resolution = 0.4 ns). We used the spectrometer as a bandpass filter of PL signal. Fitting to the time-resolved spectra was done with double exponential decay curves convolved with a function reflecting the system time response. Among the two deduced time constants, we treated the faster decay rate as the experimental value in the discussion. For the intensity correlation measurements, we added a beamsplitter and another avalanche photodiode to build a Hanbury Brown-Twiss interferometer. The $g^{(2)}[0]$ values were deduced by dividing the area of the time zero peak by the averaged area of the remaining peaks. The results in Figs. 5 were taken at 3 K using the pulse laser (831 nm, 3.8 µW).

**ACKNOWLEDGMENT**

The authors thank M. Holmes, S. Ishida, K. Watanabe, A. Osada, K. Kuruma, A. Tamada and M. Fuwa for fruitful discussions. This work was supported by JSPS KAKENHI Grant-in-Aid for Specially Promoted Research (15H05700), KAKENHI 16K06294 and a project of the New Energy and Industrial Technology Development Organization (NEDO).

22. C. P. Dietrich, A. Fiore, M. G. Thompson, M. Kamp, and S. Höfling, "GaAs integrated quantum photonics: Towards compact and multi-functional quantum photonic integrated circuits," Laser Photon. Rev. **10**, 870–894 (2016).

23. P. B. Deotare, M. W. McCutcheon, I. W. Frank, M. Khan, and M. Lončar, "High quality factor photonic crystal nanobeam cavities," Appl. Phys. Lett. **94**, 121106 (2009).

24. R. Ohta, Y. Ota, M. Nomura, N. Kumagai, S. Ishida, S. Iwamoto, and Y. Arakawa, "Strong coupling between a photonic crystal nanobeam cavity and a single quantum dot," Appl. Phys. Lett. **98**, 173104 (2011).

25. E. Kuramochi, H. Taniyama, T. Tanabe, K. Kawasaki, Y.-G. Roh, and M. Notomi, "Ultrahigh-Q one-dimensional photonic crystal nanocavities with modulated mode-gap barriers on SiO2 claddings and on air claddings.," Opt. Express **18**, 15859–15869 (2010).

26. Y. Halioua, A. Bazin, P. Monnier, T. J. Karle, G. Roelkens, I. Sagnes, R. Raj, and F. Raineri, "Hybrid III-V semiconductor/silicon nanolaser.," Opt. Express **19**, 9221–31 (2011).

27. A. Faraon, I. Fushman, D. Englund, N. Stoltz, P. Petroff, and J. Vuckovic, "Dipole induced transparency in waveguide coupled photonic crystal cavities," Opt. Express **16**, 12154 (2008).
16

# Supplementary Information

**S1. Sample Fabrication**

PhC airbridge nanobeam cavities were fabricated into a 130 nm-thick GaAs slab containing a layer of self-assembled InAs QDs grown by molecular beam epitaxy. We employed conventional nanofabrication processes including electron beam lithography and both wet and dry etching. We simply used the same process conditions as those used when studying high $Q$ III-V nanobeam cavities [1,2], which in principle could be further optimized individually.

Transfer printing was performed with a homemade printing apparatus, a photograph of which is shown in Fig S1(a). The system is composed of two movable stages and an optical microscope. The left stage (highlighted in green) holds a glass plate, on which a transparent rubber stamp is attached. The stamp is made of polydimethylsiloxane (PDMS, Sylgard184, Dow Corning). The PDMS stamp has 1-μm-thick square bumps with a side length of 30 μm for selective sample pick up. The position of the stamp can be controlled with fine adjusters in the three axes. With the top stage, the pitch and roll of the stamp can also be controlled. The right stage (blue) holds SPS and waveguide samples on the top, the positions of which can be finely tuned by the combinations of fine adjusters and piezo actuators. The sample rotation can also be corrected using an incorporated rotational stage. The sample image was obtained by the microscope, the magnification of which can be switched by rotating the turret equipping objective lenses.

For the transfer printing, first, we attached a PDMS stamp to an appropriate airbridge nanobeam cavity under the microscope, as shown in Fig. S1(b). Then, we quickly peeled the stamp off by moving an actuator in the vertical direction (Fig. S1(c)). The peeling speed is roughly 3 mm/s. The success probability of this picking up process is about 70~80% in the current setup and condition. Then, we brought the lifted nanobeam cavity onto a target waveguide. Subsequently, the cavity was carefully loaded above the waveguide manually using the piezo actuators (Fig. S1(d)). For the accurate alignment between the elements, we used cross marks patterned on them. Figure S1(e) shows a picture during the nanocavity release by slowly peeling the stamp off. A microscope image of a completed sample is shown in Fig. 3(a) in the main text. The transferred SPS is bonded tightly on the waveguide wafer via van der Waals force [3]. When integrating two nanocavities into a waveguide as shown in Fig. 5(a), we did not see significant disturbance on the printing process by the pre-located nanocavity. This suggests the possibility for dense integration of a larger number of SPSs by transfer printing, which would be required for realizing large-scale quantum PICs.

Regarding the printing accuracy, we evaluated several different printed nanocavities and deduced that the deviations of the sample positions were less than 100 nm on average. Unwanted sample rotations are found to be less than 1 degree.

In the current work, we did not pre-select a suitable nanobeam cavity. Therefore, only one out of three to four samples contain QDs resonating with the cavity mode. The sample discussed in the main text is one of such samples. For the sample with two SPSs on the single waveguide, we prepared 8 pairs of such structure. One of them have QDs in each cavity and has been used for the discussion in the main text. This randomness in the SPS fabrication can be easily avoided by pre-selecting suitable QD SPSs by optical experiments prior to the transfer processes.

We used transfer printing for the fabrication of glass-cladded wire waveguides. For this purpose, first, we prepared airbridge wire waveguides with grating exit ports [4] into a 130 nm-thick GaAs. The waveguide width was chosen to be 220 nm as discussed in the supplementary section 4. We placed the waveguides on a glass substrate by transfer printing. We then formed an upper clad on the waveguide by a spin-on-glass process (FOX15, Dow Corning). The thickness of the glass above the waveguide ($= d$) was precisely controlled to be 300 nm for the first sample shown in Fig.

3(a) by tuning the amount of solvent in the liquid glass material and the spin speed. For the experiments in Fig. 5(a), $d$ = 270 nm was used.

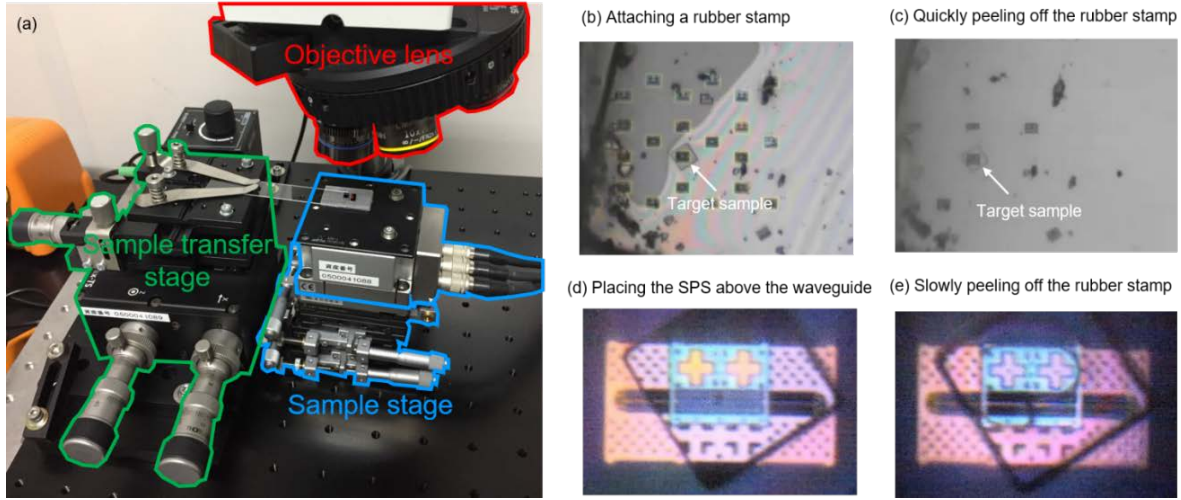

Fig. S1. Transfer printing apparatus and microscope images of each process step.

## S2. Device design

We optimized the SPS based on the PhC nanobeam cavity on a glass-cladded wire waveguide [5] so as to maximize the single photon coupling efficiency into the waveguide. First, we designed the PhC nanobeam cavity on a flat glass substrate [6]. We considered a GaAs nanobeam with a width of 370 nm and a thickness of 130 nm. The air holes were patterned with a period ($a$) of 230 nm and radii of 53 nm. The lattice period was modulated near the cavity center so as to support a very high Q-factor of over 5 million for the fundamental cavity mode at the normalized frequency ($a/\lambda$) of 0.249, which was computed using finite difference time domain (FDTD) simulations. Details of the nanocavity design are described in the supplementary section 3. Then, we simulated light coupling into the glass-cladded waveguide placed directly below the cavity with separation $d$. We set the waveguide width and thickness to be 220 nm and 130 nm, respectively. These parameters were chosen so as to maximize the cavity-waveguide coupling strength for a given $d$ (see the supplementary section 4): this optimization is essential to increase the maximum possible cavity-waveguide coupling efficiency ($\eta$) in design. The calculation of $\eta$ starts with the simulation of the investigated cavity mode until reaching to its steady state by FDTD method. Then, we measured all the light leakage from the simulator, together with that from the waveguide. In this way, we can deduce $\eta$ by the ratio of waveguide light leakage to that from the whole domain.

In an analytical fashion, we can describe $\eta$ by the following equation:

$$\eta = \frac{Q_{wg}^{-1}}{Q_0^{-1} + Q_{wg}^{-1} + Q_{scatter}^{-1}}, \tag{1}$$

where $Q_0$ and $Q_{wg}$ are the design cavity $Q$ without and with the waveguide, respectively. $1/Q_{scatter}$ expresses the additional photon loss into free space due to the introduction of the waveguide. From the equation, for realizing a high $\eta$, it is vital to design a very high $Q_0$ and a low $Q_{wg}$, while suppressing $1/Q_{scatter}$. The high $Q_0$ (>5 million) of our PhC nanobeam cavity is highly suitable for increasing $\eta$. The reduction of $1/Q_{scatter}$ is possible by taking a large enough $d$, which in turn exponentially increases $Q_{wg}$. We overcame this difficulty by optimizing the waveguide parameters so as to minimize $Q_{wg}$ to 1,300 for $d = 300$ nm (see the supplementary section 4). In this design, $1/Q_{scatter}$ becomes negligible and $\eta$ reaches to be 99.4%. In contrast, for $d$s much smaller than 300 nm, the above discussion based on the perturbation theory (coupled mode theory) does not hold anymore, since the index modulation by the waveguide becomes too strong to treat as a perturbation for the cavity mode. In this case, the significant cavity-to-free space leakage is turned on, resulting in a high $1/Q_{scatter}$ value and reduced $\eta$s, as plotted in Fig. 2(c) in the main text.

For evaluating the emitter-cavity coupling efficiency ($\beta$), we assumed that the spontaneous emission from the QD occurs into either the cavity mode or free space. The average spontaneous emission rate of QDs in unprocessed area ($\gamma$) was measured to be $\gamma = 1$ GHz. The nanocavity enables fast spontaneous emission into the cavity mode with a rate of $F_p\gamma$, where $F_p$ is the Purcell factor proportional to the value of $Q/V$ of the cavity mode. Meanwhile, the photonic bandgap effect in the one-dimensional nanobeam PhC suppresses the spontaneous emission rate of embedded QDs [2]. Consistent with the previous work, we experimentally measured $\gamma_{\text{other}}$ to be ~ $0.5\gamma$. With these parameters, $\beta$ can be calculated from the following equation:

$$\beta = \frac{F_p\gamma}{F_p\gamma + \gamma_{other}}, \qquad (2)$$

For the designed SPS with an overall cavity Q-factor of 1,300 at $d = 300$ nm, the maximum possible $F_p$ is 250, resulting in a near-unity $\beta$ of 99.7%. Overall, the total single photon coupling efficiency to the waveguide ($\eta\beta$) becomes a near-ideal value of 99.2%. Such high $\eta\beta$s are also confirmed in direct simulations of them using a dipole source emulating the single QD (see the supplementary section 5). It is noteworthy that our design strategy for near-unity $\eta\beta$ can be applied to different material systems such as those using Si and $Si_3N_4$ (see the supplementary section 6). Moreover, the design is robust against the misalignment between the cavity and the waveguide: a 200-nm cavity position deviation reduces the total efficiency by less than 1% (see the supplementary section 7). From these results, our waveguide-coupled SPS design compatible with the transfer printing process could be regarded as one of the most viable routes to introduce single photons into diverse PIC platforms, regardless of material combinations.

## S3. Details of the nanocavity design

We designed the PhC nanobeam cavity so as to possess an ultra-high Q-factor when being placed on flat glass [6] (refractive index $n = 1.45$). A schematic of the design is drawn in Fig. S2. In the design, we considered a GaAs-based ($n = 3.4$) nanobeam with a width ($w$) of 370 nm and a height ($h$) of 130 nm. The nanobeam is patterned with air holes with radii ($r$) of 59.8 nm and a period ($a$) of 230 nm. For the cavity formation, we disturbed the period of the air holes quadratically: $a_1$, $a_2$, $a_3$, $a_4$, $a_5$ and $a_6$ equals to 0.84$a$, 0.844$a$, 0.858$a$, 0.88$a$, 0.911$a$ and 0.951$a$, respectively. We assumed the thickness of the glass to be 1.5 μm in the model. We computed properties of the fundamental cavity mode resonating at a normalized frequency of 0.249 $a/\lambda$ by the FDTD method. We obtained a high Q-factor of $5.4\times10^6$ and a small mode volume of $0.434(\lambda/n)^3$.

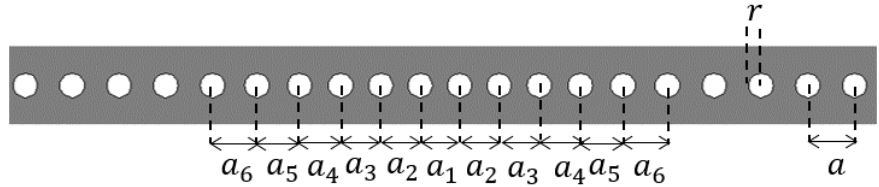

Fig. S2. Detailed schematic of the cavity design.

**S4. Maximizing the cavity-waveguide coupling strength for a given $d$**

As we discussed in the main text, it is essential for realizing high $\eta$ to achieve the maximum possible coupling between the PhC nanobeam cavity and the waveguide for a given cavity-waveguide distance, $d$. In the current work, we optimized the waveguide width so as to maximize the coupling [5]. Figure S3 shows an evolution of calculated cavity Q-factors as a function of the waveguide width for the case with $d = 300$ nm. By changing the width, cavity Q-factors increase or decrease and have its bottom around the width of 220 nm. The reduction of $Q$ predominantly stems from the cavity-waveguide coupling, rather than the cavity-to-free space leakage. Indeed, we confirmed near unity $\eta$s for the cases using waveguide widths around 220 nm by directly calculating $\eta$s through the radiation power distribution. In the current work, we set the waveguide width to be 220 nm.

We interpreted that the maximum coupling was obtained when achieving the largest overlap integral between the confined cavity mode and the propagating waveguide mode in the sense of the coupled mode theory [7]. From a different point of view, this can be understood as a result of phase matching between the two modes. Namely, by modifying the width, the waveguide dispersion was adjusted to cross the cavity frequency around the bandedge of the PhC nanobeam, the field components around which, in this case, predominantly constitute the cavity mode.

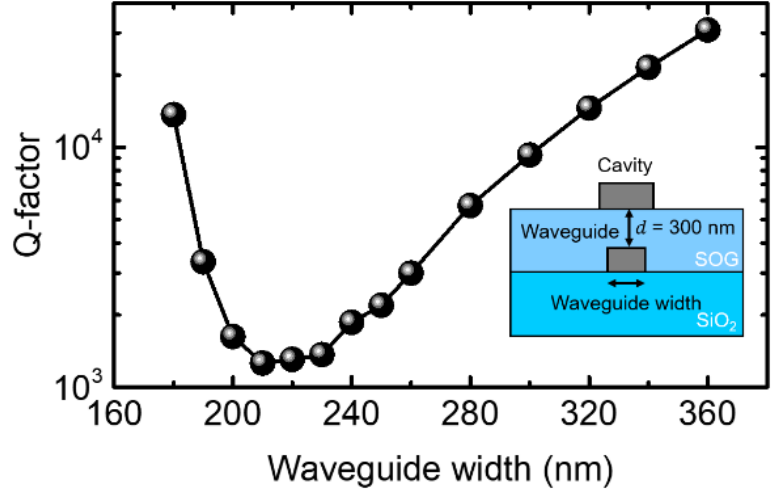

Fig. S3. Waveguide width dependence of the simulated cavity-waveguide coupling.

**S5. Direct simulation of *ηβ* by a dipole source**

For further verifying the high total single photon coupling efficiencies, *ηβ*s, we performed direct numerical simulations of them by using a point dipole as a source of the radiation [8]. In this way, we can directly simulate *ηβ* by monitoring the power distribution to the waveguide compared to the whole radiated power. Here, the dipole source was assumed to be linearly polarized, tuned to the resonance of the fundamental cavity mode and positioned to the cavity center, where the electric field of the fundamental mode is the strongest. Figure S4 shows a comparison of the directly calculated *ηβ* (red) with those by the separated calculations as discussed above (black). The two curves match very well each other, confirming the validity of the separated simulations of *η* and *β* for estimating *ηβ*. For $d = 300$ nm, the simulation using the point dipole source results in $\eta\beta$ = 99.8%. The remaining deviations between the two curves are considered to be due to the finite simulation accuracy in the FDTD method (which is limited by the spatial grid size and time length of the calculation, etc.).

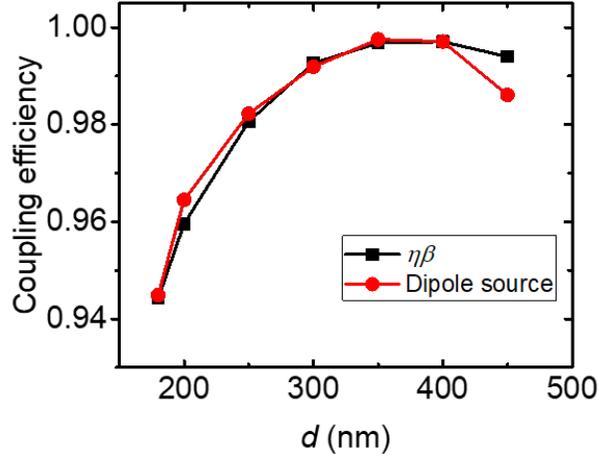

Fig. S4. Total single photon coupling efficiency calculated by a dipole source.

**S6. Simulated single photon coupling efficiencies into waveguides made of Si and $Si_3N_4$**

Our SPS design strategy is versatile and robust for the change in material platform. We examined this by considering waveguides made of Si ($n$ = 3.5) and $Si_3N_4$ ($n$ = 2.0), both of which are standard material for the fabrication of PICs.

For Si waveguide, we re-designed an InP-based ($n$ = 3.5) nanobeam cavity such that it resonates within the telecom wavelength band at 1.55 μm, where Si is transparent. A design schematic is shown in Fig. S5(a). In the cavity design, we set $a$ = 380 nm, $r$ = 99 nm, $w$ = 640 nm and $h$ = 220 nm, while employing the same modulation rule of the air hole periods described above for defining the defect cavity region. Without the waveguide, the nanobeam cavity on plane glass exhibits a very high Q-factor of $5.2 \times 10^6$ and a small $V$ of $0.463(\lambda/n)^3$. For the waveguide coupling, we assumed a Si waveguide width and thickness of 400 and 210 nm, respectively. By setting the cavity-waveguide distance *d* to be 500 nm, a near-unity $\eta$ of 99.5% was deduced based on power distributions obtained in FDTD simulations. A field profile of the simulated mode at the steady state is shown in Fig. S5(b). The cavity *Q* with the waveguide was 2,310, which results in an $\eta$ of 99.9% and well explains the simulated $\eta$. Using the calculated *Q* and *V*, a very high $\beta$ of 99.8% can be deduced. Overall, a very high single photon coupling efficiency $\eta\beta$ of 99.3% was obtained in these simulations. Meanwhile, for $Si_3N_4$ systems, we used the same nanobeam cavity design used in Fig. 2(a) in the main text. A design schematic for this structure is shown in Fig. S5(c). The $Si_3N_4$ waveguide are assumed to have a width of 1,000 nm and a thickness of 400 nm. With $d$ = 200 nm, the designed SPS exhibits a very high $\eta$ of 99.1%, together with a *Q* of 7,900. In this case, $\beta$ results in 99.9% and thereby $\eta\beta$ was deduced to be 99.0%. A computed field profile at the steady state is shown in Fig. S5(d).

These results clearly demonstrate that our design strategy is suitable for the introduction of highly efficient SPSs into diverse PIC material platforms. We emphasize that transfer printing enables the separated optimization of the fabrication processes for the SPSs and the waveguides, opening the way to high quality assembly of SPSs on PICs.

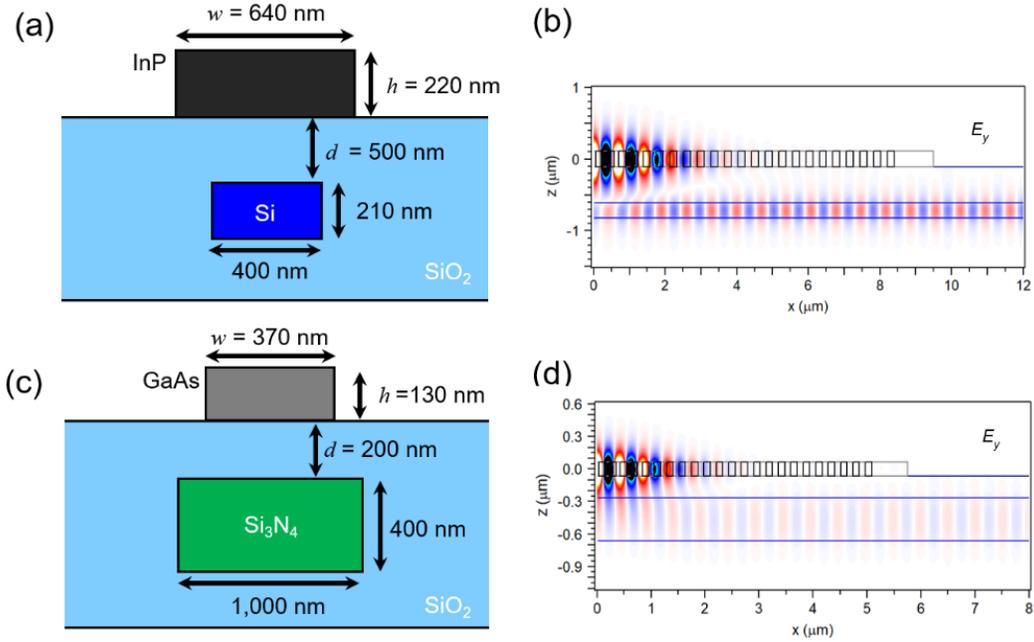

Fig. S5. Simulation of cavity-waveguide coupling efficiency for different material combinations.

**S7. Effect of misalignment of the cavity with respect to the waveguide**

The designed SPSs can keep the high $\eta$s even under the presence of misalignment of the cavity to the waveguide. First, we examined the robustness against the simple in-plane deviation, $\delta$, which is defined by the center-to-center distance between the cavity nanobeam and waveguide. Calculated Q-factors are shown in Fig. S6 (a). Up to $\delta = 200$ nm, we observed moderate Q-factors below 4,000, which is low enough to support high $\eta$s over 99%. Indeed, the low $Q$ is predominantly due to the waveguide coupling and a high $\eta$ of 99.5% for $\delta = 200$ nm was confirmed by numerically calculated power distributions by FDTD method.

Then, we checked the design robustness against the rotation, which is quantified by the angle between the cavity and the waveguide, $\theta$. Figure S6 (b) summarizes the simulated device Q-factors as a function of $\theta$. Again, we did not find a significant increase of the Q-factor, suggesting that the high $\eta$ is well maintained. For $\theta = 10$ degrees, we confirmed a high $\eta$ over 99% by the FDTD method.

We also confirmed the situations in which both finite delta and $\theta$ exist. Figure S6(c) summarizes the cases for $\delta = 100$ and 200 nm with $\theta$ from 0 to 10 degrees. Apparently, low $Q$s are still supported even under the existence of the combined misalignments. For $\delta = 200$ nm and $\theta = 10$ degrees, we confirmed a high $\eta$ of 99.1%. Assuming the experimentally observed misalignment of delta smaller than 100 nm and of $\theta$ less than 1 degree, we can conclude that the near-unity $\eta$ over 99% can be achieved even by the current transfer printing technology, if we can experimentally achieve $Q_0$ over 50,000.

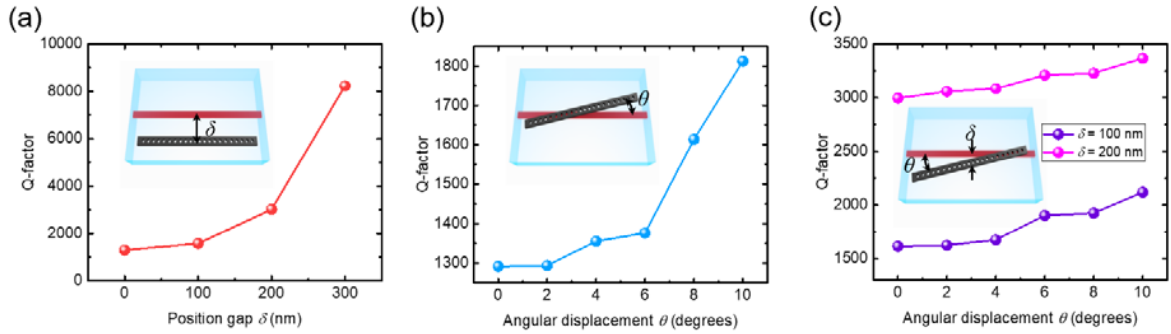

Fig. S6. Calculated Q-factors under the presence of the misalignment.

**S8. Single photon transport under the presence of two cavities on a waveguide**

We simulated light transport for the case that the two cavities are simultaneously integrated onto an individual waveguide. We used two different PhC nanobeam cavities: they differ in $a$ (230 nm and 235 nm), while share the same $r/a = 0.26$, $w = 370$ nm and $h = 130$ nm. The two cavities resonate at 924 nm for $a = 230$ nm and 937 nm for $a = 235$ nm in the simulation. For investigating the light propagation, we selectively excite one of the two cavities and computed its evolution until reaching to the steady state in our numerical simulator. Figures S7(a) and (b) show the calculated field profiles at the steady states. It is clearly seen that the light transport did not significantly disturbed even with the presence of the other cavity. By monitoring the power distribution into the waveguide, we deduced $\eta$s over 99% for the waveguide coupling from the two cavities. The minimal disturbance on the light transport is largely due to the large-enough frequency detuning between the two cavity modes.

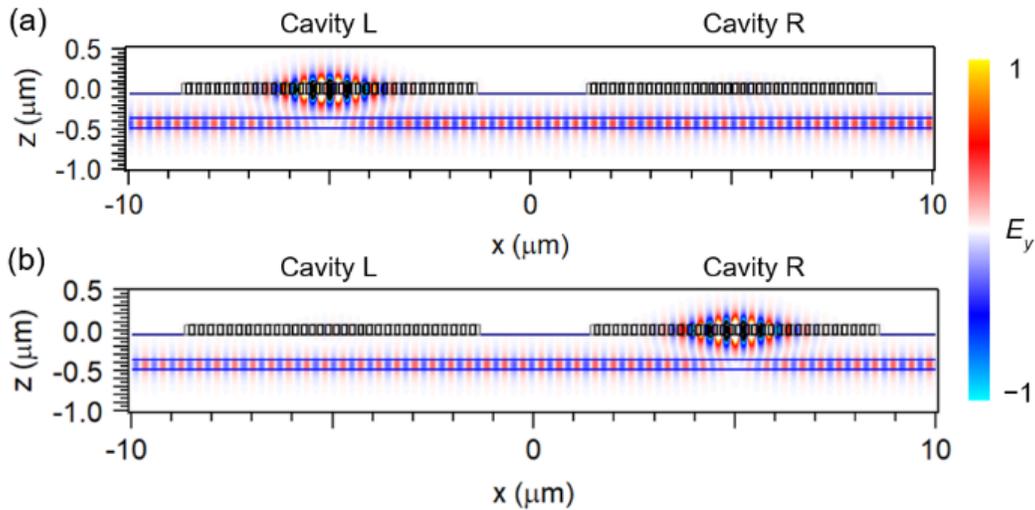

Fig. S7. Simulated light transport in the waveguide under the presence of two cavities.

## S9. Experimental estimation of $\eta$ and $\beta$.

The estimation of experimental cavity-waveguide coupling efficiencies ($\eta_{exp}$) was done based on the measured cavity Q-factors. First, we measured several nanobeam cavities that were placed on flat glass and did not coupled to the waveguide. By fitting the measured PL peaks with Lorentzian function, we deduced an average cavity Q-factor on flat glass ($Q_{ave}$) to be 13,000. In addition, the emission spectrum measured for the cavity mode coupled to the waveguide was fitted to deduce the experimental Q-factor ($Q_{exp}$) of 3,600. The larger $Q_{exp}$ than the designed $Q$ of 1,300 for $d$ = 300 nm can be explained by the deviation of the structural parameters in the fabricated device: in particular, the change in the cavity resonance wavelength disturbs the phase matching between the cavity and waveguide and degrades their coupling strength. Using the measured cavity Q-factors, we deduced an $\eta_{exp}$ of 72% based on the following equation:

$$\eta_{exp} = \frac{Q_{exp}^{-1} - Q_{ave}^{-1}}{Q_{exp}^{-1}}. \tag{3}$$

This equation assumes that the observed reduction of the Q-factor when introducing the waveguide is dominated by the leakage of photons into the waveguide. This assumption is fairly reasonable in the current situation where $d$ is large enough to suppress the additional cavity photon leakage into free space, as confirmed in the numerical simulations (see the supplementary section 2). For the case when loading the two cavities to the single waveguide, we measured a Q-factor of 1,000 (950) for the left (right) cavity, resulting in $\eta_{exp}$ of 92% (93%). In this case, $\eta_{exp}$ improved due to the slight reduction of $d$ to 270 nm, which reduces the cavity Q-factors and increases the coupling between the cavities and the waveguide.

Regarding the estimation of the experimental $\beta$ ($\beta_{exp}$), we performed time-resolved PL measurements. First, we measured emission decay rates of several single QDs that were embedded in PhC nanobeams on plane glass and decoupled from any cavity modes. The average decay rate was measured to be 0.5 GHz (= $\gamma_{other}$), which is roughly half of that for QDs in an unprocessed region of the sample. This reduction of the decay rate stems from partial photonic bandgap effect in the PhC nanobeam [2]. Then, we measured emission decay rates of the investigated QD emission peaks coupled to the cavity mode ($\gamma_{exp}$). With these decay rate values, we deduced $\beta_{exp}$ by the following equation [9]:

$$\beta_{exp} = \frac{\gamma_{exp} - \gamma_{other}}{\gamma_{exp}}. \tag{4}$$

For the QD discussed in Fig. 4(b) in the main text, $\gamma_{exp}$ was measured to be 3.8 GHz at the resonance, resulting in $\beta_{exp}$ of 87%. For the QD in the left (right) cavity in Fig. 5(d)((e)) in the main text, $\gamma_{exp}$ was measured to be 2.5 (1.1) GHz at 3 K, at which the QD is detuned from the cavity resonance by 2.0 (1.8) nm. The resulting $\beta_{exp}$ was 80% (57%). The experimental single photon coupling efficiencies into the waveguide were simply obtained by multiplying the two efficiencies, that is $\eta_{exp}\beta_{exp}$.

**S10. Achievable single photon coupling efficiency with current technology**

In the current demonstration, $\beta_{exp}$ did not reach to the maximum possible value probably due to the deviation of the QD position from the cavity field maximum, which degrades the Purcell effect enhancement. If the QD position was optimum, the maximum $\beta_{exp}$ would reach to 99.7%. Moreover, by optimizing the waveguide-cavity distance (*d*) to 250 nm, it could be possible to increase $\eta$ up to 96.3%, with slight reduction of the maximum possible $\beta$ to be 99.4%. Overall, it would be possible to achieve a total single photon coupling efficiency of $\eta\beta = 95.7\%$ even under the present quality of the nanocavity fabrication. Meanwhile, we have already demonstrated a cavity Q-factor over 50,000 for PhC nanocavities [10]. With this fabrication quality, it would be possible to achieve $\eta\beta$ over 98.4% with $d = 250$ nm. Moreover, if the Q-factor reached to that of the state-of-the-art PhC nanobeam cavity (~ one million), the maximum possible $\eta\beta$ would become 99.6% for $d = 300$ nm. These estimations imply that the near-unity coupling of single photons into the waveguide is already within reach of the current process technology. It is noteworthy that the optimum *d* for realizing the highest possible $\eta\beta$ varies with the achievable cavity Q-factor when not coupled to the waveguide, since $\eta\beta$ is determined between the waveguide coupling and the Purcell effect which increase or decrease depending on the total Q-factor.

**S11. Measured evolution of spontaneous emission rates as a function of detuning**

Figure S8 shows measured spontaneous emission rates plotted as a function of cavity-QD detuning for the SPS discussed in Fig. 4(b) in the main text. The fastest emission rate of 3.8 ns$^{-1}$ is achieved at the emitter-cavity resonance. The emission rates become slower when detuned from the resonance. These behaviors support our conclusion that the emission rate enhancement originates from the Purcell effect within the nanocavity. The solid line in the figure shows a Lorentzian peak with the same linewidth of the cavity mode ($Q$ = 3,600). The emission rate evolution does not match with the Lorentzian peak, being apart from the expectation from the conventional theory of the Purcell effect in cavity. We consider that this discrepancy is due to coupling of the QD to acoustic phonons. The phonon-assisted Purcell effect in QDs is known to support a broad range of the emission rate enhancement [11]. Indeed, the widened rate enhancement curve with a full width half maximum of 1.5 nm (2.3 meV) reasonably match with those discussed in the literature.

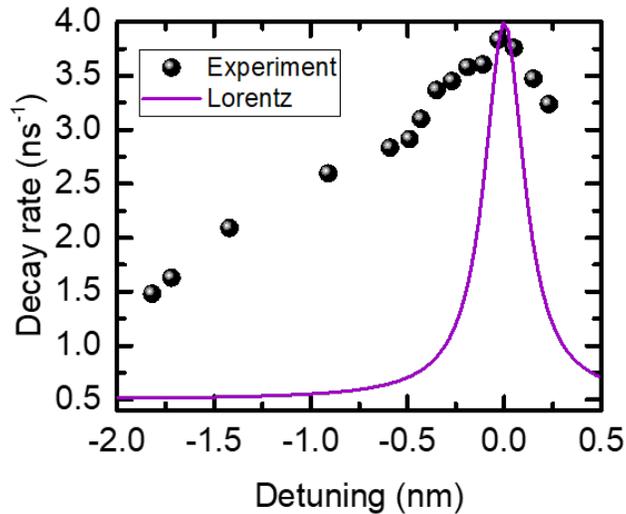

Fig. S8. Detuning dependence of the decay rates of the QD emission.

**References**

1. R. Ohta, Y. Ota, M. Nomura, N. Kumagai, S. Ishida, S. Iwamoto, and Y. Arakawa, "Strong coupling between a photonic crystal nanobeam cavity and a single quantum dot," Appl. Phys. Lett. **98**, 173104 (2011).

2. Y. Ota, R. Ohta, N. Kumagai, S. Iwamoto, and Y. Arakawa, "Vacuum Rabi Spectra of a Single Quantum Emitter," Phys. Rev. Lett. **114**, 143603 (2015).

3. J. Justice, C. Bower, M. Meitl, M. B. Mooney, M. a. Gubbins, and B. Corbett, "Wafer-scale integration of group III–V lasers on silicon using transfer printing of epitaxial layers," Nat. Photonics **6**, 612–616 (2012).

4. A. Faraon, I. Fushman, D. Englund, N. Stoltz, P. Petroff, and J. Vuckovic, "Dipole induced transparency in waveguide coupled photonic crystal cavities," Opt. Express **16**, 12154 (2008).

5. Y. Halioua, A. Bazin, P. Monnier, T. J. Karle, G. Roelkens, I. Sagnes, R. Raj, and F. Raineri, "Hybrid III-V semiconductor/silicon nanolaser.," Opt. Express **19**, 9221–31 (2011).

6. E. Kuramochi, H. Taniyama, T. Tanabe, K. Kawasaki, Y.-G. Roh, and M. Notomi, "Ultrahigh-Q one-dimensional photonic crystal nanocavities with modulated mode-gap barriers on SiO2 claddings and on air claddings.," Opt. Express **18**, 15859–15869 (2010).

7. E. Waks and J. Vuckovic, "Coupled mode theory for photonic crystal cavity-waveguide interaction.," Opt. Express **13**, 5064–73 (2005).

8. Y. Xu, J. Vučković, R. Lee, O. Painter, A. Scherer, and A. Yariv, "Finite-difference time-domain calculation of spontaneous emission lifetime in a microcavity," JOSA B **16**, 465–474 (1999).

9. M. Arcari, I. Söllner, A. Javadi, S. Lindskov Hansen, S. Mahmoodian, J. Liu, H. Thyrrestrup, E. H. Lee, J. D. Song, S. Stobbe, and P. Lodahl, "Near-Unity Coupling Efficiency of a Quantum Emitter to a Photonic Crystal Waveguide," Phys. Rev. Lett. **113**, 93603 (2014).

10. Y. Ota, S. Iwamoto, N. Kumagai, and Y. Arakawa, "Spontaneous Two-Photon Emission from a Single Quantum Dot," Phys. Rev. Lett. **107**, 233602 (2011).

11. U. Hohenester, A. Laucht, M. Kaniber, N. Hauke, A. Neumann, A. Mohtashami, M. Seliger, M. Bichler, and J. J. Finley, "Phonon-assisted transitions from quantum dot excitons to cavity photons," Phys. Rev. B **80**, 201311 (2009).
33